\documentclass[runningheads]{llncs}
\usepackage[T1]{fontenc}
\usepackage{graphicx}
\usepackage{booktabs}
\usepackage{multirow}
\usepackage{todonotes}
\usepackage{subcaption}
\begin{document}
\title{Few-shot learning for security bug report identification}
\author{Muhammad Laiq} 
%\and
%Second Author\inst{2,3}\orcidID{1111-2222-3333-4444} \and
%Third Author\inst{3}\orcidID{2222--3333-4444-5555}}
\institute{Department of Communication, Quality Management and Information Systems \\
Mid Sweden University, Campus Östersund, Sweden \\
\email{muhammad.laiq@miun.se}}
\maketitle              % typeset the header of the contribution
\begin{abstract}
%The abstract should briefly summarize the contents of the paper in
Security bug reports require prompt identification to minimize the window of vulnerability in software systems. Traditional machine learning (ML) techniques for classifying bug reports to identify security bug reports rely heavily on large amounts of labeled data. However, datasets for security bug reports are often scarce in practice, leading to poor model performance and limited applicability in real-world settings.
In this study, we propose a few-shot learning-based technique to effectively identify security bug reports using limited labeled data.
We employ SetFit, a state-of-the-art few-shot learning framework that combines sentence transformers with contrastive learning and parameter-efficient fine-tuning. The model is trained on a small labeled dataset of bug reports and is evaluated on its ability to classify these reports as either security-related or non-security-related.
Our approach achieves an AUC of 0.865, at best, outperforming traditional ML techniques (baselines) for all of the evaluated datasets. This highlights the potential of SetFit to effectively identify security bug reports. SetFit-based few-shot learning offers a promising alternative to traditional ML techniques to identify security bug reports. The approach enables efficient model development with minimal annotation effort, making it highly suitable for scenarios where labeled data is scarce.

\keywords{Few-shot learning \and SetFit \and Issue classification \and Bug report classification \and Security bug reports.}
\end{abstract}

\section{Introduction}\label{sec:Introduction}
Software projects receive a large number of bug reports through bug tracking systems such as JIRA\footnote{https://www.atlassian.com/software/jira/templates/bug-tracking}, Bugzilla\footnote{https://www.bugzilla.org/}, and GitHub Issues\footnote{https://github.com/features/issues}. These reports vary in nature, from enhancement requests to usability issues, performance complaints, and security vulnerabilities. Among the various types of bug reports, security-related bug reports are particularly critical because they may expose the system to exploitation. Failure to identify and address such vulnerabilities in a timely manner can have severe consequences. Manually reviewing bug reports to identify security-related bug reports is time-consuming, error-prone, and not scalable for large-scale projects that may receive thousands of bug reports \cite{wu2021data}. 

Several automated techniques have been proposed in the literature (e.g., \cite{peters2017text,zhou2017automated,goseva2018identification,shu2019better}) to assist in identifying security bug reports, particularly those based on traditional machine learning (ML) techniques. These techniques use supervised ML models trained on historical bug reports to predict whether a new report is security-related. Traditional ML techniques come with significant limitations. In particular, they require a large amount of labeled training data, which is rarely available for security bug reports in practice \cite{ohira2015dataset,francca2025gpts}. Moreover, the class imbalance between security and non-security reports further impacts the models' performance, often leading to biased or unreliable predictions.

To overcome data scarcity limitations in security bug reports, recent work \cite{francca2025gpts} has explored the use of large language models, particularly Generative Pre-trained Transformer (GPT) models, for classifying bug reports to identify security bug reports. Although GPT-based models have shown promising results in many tasks of natural language processing/understanding, their performance in identifying security bug reports is lower than that of traditional ML techniques \cite{francca2025gpts}.

We build on these efforts by exploring an approach that is tailored to address the challenge of limited labeled data through few-shot learning. In particular, we propose the use of SetFit (Sentence Transformer Finetuning) \cite{koch2015siamese} to identify security bug reports. SetFit is the state-of-the-art few-shot learning framework that combines sentence transformers with contrastive learning and parameter-efficient fine-tuning. SetFit has shown effective performance for similar tasks \cite{colavito2023few} in scenarios with limited labeled data.

This paper contributes by investigating the application of SetFit, a few-shot learning approach, to the task of identifying security bug reports by classifying bug reports. We empirically evaluate the effectiveness of SetFit on the four widely used datasets for security bug reports (i.e., Ambari, Camel, Wickey, and Derby). Our findings indicate that SetFit can achieve an AUC of 0.865, at best, outperforming traditional ML techniques (baselines) for all of the evaluated datasets \cite{francca2025gpts}.

The remainder of this paper is organized as follows. Section \ref{sec:Method} presents the research method of the study. Section \ref{sec:Results} presents the results of the study. Section \ref{sec:Discussion} discusses the findings of the study and describes the related work on identifying security bug reports. Section \ref{sec:validitythreats} describes the validity threats to the study, and Section \ref{sec:Conclusions} concludes the paper.

\section{Research Method}\label{sec:Method}

This study aims to evaluate the effectiveness of a few-shot learning approach (i.e., SetFit) in identifying security bug reports. To achieve this goal, we posed the following overarching research question: \textbf{\textit{How does SetFit perform to identify security bug reports?}} To answer the research question, we perform comparative experiments \cite{alpaydin2020introduction,wohlin2012experimentation}. 

In the following subsections, we describe the main decisions taken in the design of the experiments, including the selection of automatic techniques, the selection of datasets, the choice of the evaluation approach, and the choice of performance evaluation metrics.

\subsection{Selection of techniques}\label{sec:techniquesselection}

As baselines, we selected three traditional supervised ML techniques that were used as baselines in previous work on security bug report identification \cite{zheng2021comparative}: Support Vector Machines, Logistic Regression, and Random Forest. In addition, we introduce SetFit \cite{koch2015siamese} as a new baseline for identifying security bug reports. We will compare the performance of our proposed technique (SetFit) against these three traditional ML techniques.

\subsection{Selection of datasets}\label{sec:datasets}

To evaluate both the proposed SetFit and traditional ML techniques, we selected commonly used datasets for security bug report identification: Ambari, Camel, Derby, and Wicket. These datasets have been used as benchmarks in several prior studies~\cite{alqahtani2024security,yokoyama2024identifying,wu2021data}, including the recent work by Francca et al.~\cite{francca2025gpts}.

Table~\ref{tab:datasets} provides an overview of the datasets used in our study. All four are open-source projects, originally compiled and provided by Wu et al.~\cite{wu2021data}. Specifically, Ambari\footnote{\url{https://ambari.apache.org/}} is a Hadoop management platform, Camel\footnote{\url{http://camel.apache.org/}} is an integration framework, Derby\footnote{\url{https://db.apache.org/derby/}} is a relational database management system, and Wicket\footnote{\url{https://wicket.apache.org/}} is a Java web application framework.

\begin{table}[ht]
    \centering
\caption{Number of security bug reports and the total number of bug reports (BRs) for each selected dataset.}
%\vspace{-2mm}
    \label{tab:datasets}
    \begin{tabular}{p{2cm}p{3cm}c}
    \toprule 
\textbf{Project} &  \textbf{Security BRs}  &  \textbf{Total BRs} 
      \\\midrule

Camel &  58 (10\%) &  580 \\

Ambari & 48 (5.5\%) &  871 \\

Derby & 157 (21.5\%) & 731 \\

Wicket & 43 (6.5\%) &  663 \\
\bottomrule

\end{tabular}
\end{table}

\subsection{Evaluation approach}\label{sec:evaluationaproach} 

Several evaluation approaches exist for evaluating the performance of ML techniques, including cross-validation, testing on fixed datasets, and time-based or incremental splitting. This study uses a five-fold cross-validation approach for both traditional ML techniques and SetFit. This strategy helps reduce potential bias compared to testing on fixed datasets and improves the reliability of the results \cite{witten2002data}. To perform the 5-fold cross-validation, we divided our datasets into five equal folds. We trained the models on four of these folds for each iteration and tested them on the remaining single fold. This process was repeated five times, allowing different test sets to be used in each iteration. Finally, we averaged the results across all five folds for each technique.

\subsection{Performance evaluation metrics}\label{sec:evaluationmetrics}

The four datasets (see Table \ref{tab:datasets}) used in this study are imbalanced. Thus, to evaluate the performance of the selected techniques on imbalanced data, we used metrics considered robust to
data imbalance, i.e., Area Under the Curve (AUC) of the receiver operator characteristic (ROC) \cite{huang2005using} and Matthew’s Correlation Coefficient (MCC) \cite{matthews1975comparison}. The AUC value is determined by plotting the ROC curve, which uses the True Positive Rate (TPR) and False Positive Rate (FPR). 

\begin{equation}
TPR= \frac{TP}{TP+FN}
\end{equation}

\begin{equation}
FPR= \frac{FP}{FP+TN}
\end{equation}

MCC measures the correlation between predicted and actual values derived from the confusion matrix and can be calculated from it. MCC values range from 1 to -1, with 1 indicating perfect classification and -1 indicating perfect misclassification. An MCC of 0 means the model performs no better than random chance. In addition to the underlying metrics, AUC and MCC, we also measure Precision, Recall, and F-score.

\begin{equation}
MCC = \frac{TP \cdot TN - FP\cdot FN}{\sqrt{(TP+FP)(TP+FN)(TN+FP)(TN+FN)}}
\end{equation}

\begin{equation}
Recall = \frac{TP}{TP+FN}
\end{equation}

\begin{equation}
Precision = \frac{TP}{TP+FP}
\end{equation}

\begin{equation}
F-score = 2 * \frac{Precision * Recall}{Precision + Recall}
\end{equation}

\subsection{Data preprocessing}\label{sec:datapreprocessing}

Similar to the previous work \cite{francca2025gpts}, this work uses the description of bug reports for training models to identify security bug reports. We apply the standard data pre-processing approach to the description of bug reports, including removing special characters, numbers, and links. 

For classical ML techniques, we used Term Frequency-Inverse Document Frequency (TF-IDF) to convert textual data into sparse matrices \cite{vectorization}. For the SetFit model, we used allmpnet-base-v2, one of the top-performing pre-trained model
for embeddings \cite{colavito2023few}.

\subsection{Implementation details}\label{sec:implementationdetails}

We utilized the sklearn library \cite{scikit-lib} to apply the chosen traditional ML techniques (Support Vector Machines, Logistic Regression, and Random Forest). The optimal parameters for these techniques are identified through grid search using the sklearn library. At first, we tune the parameters of each selected traditional ML technique using grid search, and afterward, we build a model employing the best parameters.

For SetFit, we follow the official documentation from Hugging Face\footnote {https://huggingface.co/docs/setfit/index} to create a model using sentence-transformers/all-mpnet-base-v2. 

To perform our experiments using classical ML techniques and SetFit, we used a machine equipped with a Windows 11 64-bit operating system, Intel (R) Core (TM) Ultra 9, 16 cores, 64 GB RAM, and NVIDIA GeForce RTX 4060 laptop GPU, 8 GB.

\section{Results and analysis}\label{sec:Results}
To answer the posed research question (i.e., \textbf{\textit{How does SetFit perform to identify security bug reports?}}), we evaluated the performance of the selected baseline techniques, Support Vector Machines, Logistic Regression, and Random Forest, and our proposed technique (i.e., SetFit).
We evaluated these techniques using four datasets of security bug reports (i.e., Camel, Ambari, Derby, and Wicket, see Table \ref{tab:datasets}). In the following, we present the evaluation results of these techniques.

\begin{table}[!ht]
    \centering
    \small
\caption{Performance of Logistic Regression (LR), Support Vector Machines (SVM), Random Forest (RF), and SetFit for identifying security bug reports. Area under the curve (AUC) and Matthew’s Correlation Coefficient (MCC).}
%\vspace{-2mm}
    \label{tab:evaluation-results}
    \begin{tabular}{p{1.5cm}p{2.5cm}p{1.5cm}p{1.5cm}p{1.5cm}p{1.7cm}p{1.3cm}}
    \toprule 
\textbf{Dataset} &  \textbf{Technique}  &  \textbf{AUC} &  \textbf{MCC}  &  \textbf{F-Score} &  \textbf{Precision} &  \textbf{Recall} 
      \\\midrule

\multirow{4}{*}{\centering Camel}
& LR (Baseline) & 0.772 &  0.688 & 0.685 & \textbf{0.921} & 0.550 \\

&  SVM (Baseline) & 0.567 & 0.277 & 0.224 & 0.667 & 0.138 \\

&  RF (Baseline) & 0.508 & 0.055 &  0.031 & 0.200 &  0.017 \\

& SetFit (Our) &  \textbf{0.807} & \textbf{0.733} & \textbf{0.737} & 0.918 & \textbf{0.620} \\
\midrule

\multirow{4}{*}{\centering Ambari}
& LR (Baseline) & 0.500 & 0.000 & 0.000 & 0.000 & 0.000 \\

&  SVM (Baseline) & 0.500 & 0.000 & 0.000 & 0.000 & 0.000 \\

&  RF (Baseline) & 0.500 & 0.000 & 0.000 & 0.000 & 0.000 \\

& SetFit (Our) & \textbf{0.567} & \textbf{0.182} &  \textbf{0.184} & \textbf{0.310} &  \textbf{0.149} \\
\midrule

\multirow{4}{*}{\centering Derby}
& LR (Baseline) & 0.775 & 0.664 & 0.697 &  0.909 & 0.566 \\

&  SVM (Baseline) & 0.659 & 0.477 & 0.478 & 0.874 & 0.331 \\

&  RF (Baseline) & 0.641 & 0.448 & 0.438 & 0.876 & 0.293 \\

& SetFit (Our) & \textbf{0.834} &  \textbf{0.793} & \textbf{0.783} &  \textbf{0.978} & \textbf{0.669} \\
\midrule

\multirow{4}{*}{\centering Wicket}
& LR (Baseline) & 0.777 & 0.716 & 0.701 & \textbf{0.960} & 0.556 \\

&  SVM (Baseline) & 0.557 & 0.249 & 0.194 & 0.600 & 0.117 \\

&  RF (Baseline) &  0.500 & 0.000 & 0.000 & 0.000 & 0.000 \\

& SetFit (Our) & \textbf{0.865} & \textbf{0.755} & \textbf{0.804} & 0.844 & \textbf{0.770} \\
\bottomrule

\end{tabular}
\end{table}

\begin{figure}[!ht]
    \centering
    \begin{subfigure}[b]{0.45\textwidth}
        \includegraphics[width=0.98\textwidth]{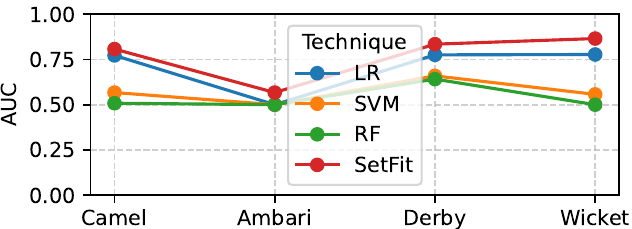}
        \caption{AUC}
        \end{subfigure}
    \quad
    \begin{subfigure}[b]{0.45\textwidth}
        \includegraphics[width=0.98\textwidth]{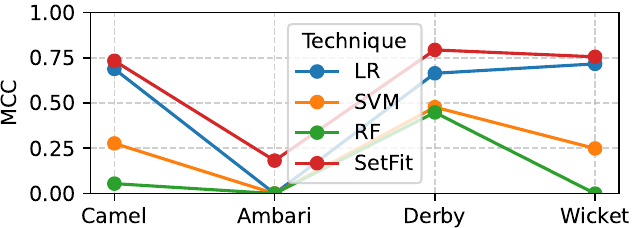}
        \caption{MCC}
     \end{subfigure}
     
    \vspace{1em} % Add vertical space between the rows

    \begin{subfigure}[b]{0.45\textwidth}
        \includegraphics[width=0.98\textwidth]{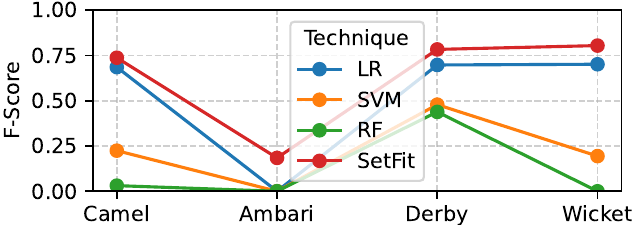}
        \caption{F-score}
    \end{subfigure}
    \quad
    \begin{subfigure}[b]{0.45\textwidth}
        \includegraphics[width=0.98\textwidth]{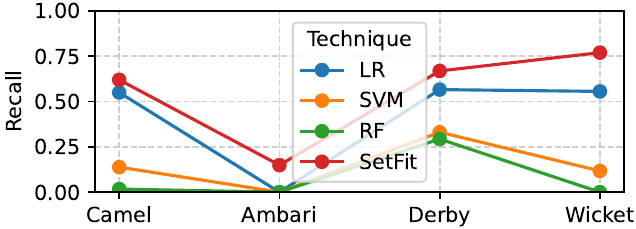}
        \caption{Recall}
    \end{subfigure}
        \begin{subfigure}[b]{0.45\textwidth}
        \includegraphics[width=0.98\textwidth]{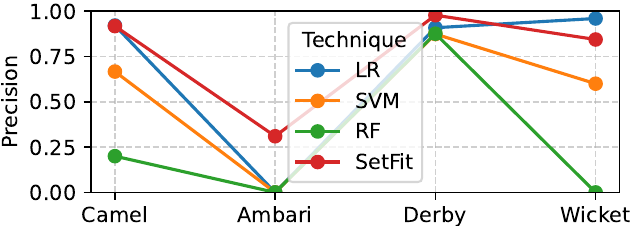}
        \caption{Precision}
        \end{subfigure}

   \caption{AUC, MCC, F-score, Recall, and Precision of the evaluated techniques (Logistic Regression (LR), Support Vector Machines (SVM), Random Forest (RF), and SetFit) for identifying security bug reports.}
    \label{fig:evaluation-results}
\end{figure}

Table \ref{tab:evaluation-results} and Figure \ref{fig:evaluation-results} present the results of our study. Across all datasets, SetFit consistently outperforms baseline techniques in nearly all metrics.

\textbf{\textit{Camel project:}} In the Camel dataset, SetFit achieves the highest scores, with an AUC of 0.807 and an MCC of 0.733. This significantly outperforms the strongest baseline, LR, with an AUC of 0.772 and an MCC of 0.688. Additionally, SetFit produces the highest F-score (0.737) and Recall (0.620), indicating its ability to identify security-relevant bug reports correctly. Unlike the baselines, these results suggest that SetFit effectively balances Precision and Recall.

\textbf{\textit{Ambari project:}} The Ambari dataset presents a particularly challenging scenario in which all baseline models fail completely, resulting in AUC and MCC scores of 0.500 and 0.000, respectively. SetFit, although not achieving high performance (AUC = 0.567, MCC = 0.182), is the only technique that is able to identify some sort of patterns from the data. These results for the Ambari dataset are consistent with previous research \cite{francca2025gpts} on identifying security bug reports.

\textbf{\textit{Derby project:}} In the Derby dataset, SetFit outperforms baselines for all metrics. It achieves the highest AUC score of 0.834 and a maximum MCC of 0.793. In addition, SetFit shows a significant improvement in Recall, with a score of 0.669 compared to LR (0.566). It also maintains a superior Precision of 0.978. These improvements result in a notably higher F-score of 0.783, demonstrating that SetFit delivers better overall classification performance.

\textbf{\textit{Wicket project:}} The results for the Wicket project further indicate SetFit’s effectiveness in identifying security bug reports. It achieves the best AUC (0.865), MCC (0.755), and F-score (0.804). Notably, SetFit surpasses the strong baseline LR in Recall (0.770 vs. 0.556) while maintaining high Precision (0.844). This balance is useful, especially when missing a security bug can be costly.

In summary, the evaluation results (see Table \ref{tab:evaluation-results} and Figure \ref{fig:evaluation-results}) show that SetFit is effective at identifying security bug reports compared to the baselines, even when there is minimal labeled data available. Its consistent performance across various datasets indicates that the SetFit-based few-shot learning approach is a promising approach for this task. Traditional baselines tend to struggle, especially in scenarios with limited data, while SetFit’s ability to generalize from just a few examples proves particularly advantageous.

\section{Discussion}\label{sec:Discussion}

This section discusses our findings and the related work on identifying security bug reports.

Bug reports may be categorized using various criteria, including their validity, priority level, or whether they pertain to security or non-security concerns. As a result, numerous studies have investigated different aspects of bug report classification \cite{laiq2025automatic}. For instance, prior research has addressed classification based on report validity \cite{laiq2022early,laiq2025comparative} as well as the identification of different issue types, such as feature requests, user questions, and documentation-related tasks \cite{laiq2023intelligent,kallis2019ticket}. In addition, other work has explored the grouping or clustering of bug reports to uncover common root causes \cite{laiq2023data,rahman2020some}.
In contrast, the present study focuses on classifying bug reports to distinguish security-related reports from those that are not.

Several studies have investigated the use of automated techniques for classifying bug reports to identify security bug reports, for example, \cite{peters2017text,zhou2017automated,goseva2018identification,shu2019better}. Among the automated techniques evaluated, traditional ML techniques are predominant, including Logistic Regression, Support Vector Machines, Random Forest, K-nearest Neighbor, and Naive Bayes. More recently, Francca et al. \cite{francca2025gpts} investigated the use of large language models (LLMs), specifically GPT-based models, for the classification of bug reports to identify security bug reports. Despite the growing popularity and general-purpose capabilities of these models, their study \cite{francca2025gpts} revealed that GPT-based models underperform compared to traditional ML techniques in this context.

In related work on bug report classification \cite{laiq2025automatic,laiq2025comparative,antoniol2008bug,herzig2013s}, where reports are categorized into bugs, questions, features, and documentation, Colavito et al. \cite{colavito2023few} applied a few-shot learning approach using SetFit \cite{koch2015siamese}. They trained their model using 200 annotated bug reports and evaluated it on another 200, achieving an F1-micro score of 0.832. Their findings demonstrate that SetFit can deliver competitive performance even with limited training data, solving the unavailability of large labeled datasets,  often required for traditional ML techniques.
Inspired by their work \cite{colavito2023few}, in this work, we applied SetFit for identifying security-related bug reports. 

Through a comparative analysis of traditional ML baselines \cite{francca2025gpts} (Logistic Regression, SVM, and Random Forest) and our proposed approach (i.e., SetFit), we demonstrate that SetFit consistently outperforms these techniques across the evaluated datasets, see details in Table \ref{tab:evaluation-results} and Figure \ref{fig:evaluation-results}.

Our findings suggest that SetFit provides a promising solution for identifying security-related bug reports. Unlike traditional ML techniques, which generally require substantial labeled datasets to perform well, SetFit performs effectively even when trained on a limited number of labeled examples. This is particularly valuable in the context of identifying security bug reports, where annotated data is both scarce and costly to obtain due to the need for domain expertise.

\section{Threats to validity}\label{sec:validitythreats}

\paragraph{\textbf{\textit{Generalizability of results:}}} In this study, we used more than one dataset (i.e., Camel, Ambari, Derby, Wicket, see details in Section \ref{sec:datasets}) to improve the generalizability of our results. Although these datasets are from different projects, they may not fully represent all software development contexts, particularly proprietary projects.

\paragraph{\textbf{\textit{Validity of the evaluation approach:}}} Different evaluation approaches, such as cross-validation and testing on a fixed set, can be used to evaluate the performance of ML techniques. To minimize potential bias in the evaluation process and enhance the reliability of the results, we utilized a 5-fold cross-validation approach (see details in Section \ref{sec:evaluationaproach}). This strategy helps to reduce experimental bias and strengthens the reliability of the outcomes \cite{witten2002data}.

\paragraph{\textbf{\textit{Validity of the evaluation metrics:}}} The datasets (i.e., Camel, Ambari, Derby, Wicket, see details in Section \ref{sec:datasets}) used in this study are imbalanced. To reduce potential bias in interpreting the results and improve the results’ reliability, we used the evaluation metrics that are considered robust to data imbalance, i.e., AUC \cite{huang2005using} and MCC \cite{matthews1975comparison}.

\paragraph{\textbf{\textit{Reliability of training data:}}} The data used to train the techniques may pose validity threats. For instance, bug reports might be misclassified. To minimize this risk, we used four widely used and manually validated datasets \cite{wu2021data} to evaluate our proposed technique to identify security bug reports.

\section{Conclusions and future work}\label{sec:Conclusions}

This paper investigates the use of SetFit, a state-of-the-art few-shot learning approach, to identify security bug reports. At best, our approach achieves an AUC of 0.865, outperforming traditional ML baselines (Logistic Regression, SVM, and Random Forest) for all of the evaluated datasets. The findings emphasize the significance of utilizing pre-trained language representations and contrastive learning to improve the identification of security bug reports. This also showcases the effectiveness of SetFit for this task.

In conclusion, SetFit-based few-shot learning presents a promising alternative to traditional ML techniques for identifying security bug reports. This approach facilitates efficient model development with minimal annotation effort, making it particularly suitable for situations where labeled data is limited.

Future work will focus on extending the evaluation of the proposed technique using a broader range of projects, incorporating data from both open-source and closed-source contexts. Additionally, we will investigate the explainability of the model's predictions when identifying security bug reports using SetFit.

\begin{credits}

\subsubsection{\discintname}
The authors declare that they have no known competing financial interests or personal relationships that could have appeared to influence the work reported in this paper.
\end{credits}

\bibliographystyle{splncs04}
\bibliography{paper}

\end{document}